\documentclass[pre,twocolumn,groupedaddress]{revtex4}
\usepackage{amsmath}
\usepackage{amsfonts}
\usepackage{amssymb}
\usepackage{graphicx}
\usepackage{ multirow}

\DeclareMathOperator{\sgn}{sgn}

\begin{document}
\title{Charge constrained density functional molecular dynamics for simulation
of condensed phase electron transfer reactions}
\author{Harald Oberhofer}

\author{Jochen Blumberger}
\email{jb376@cam.ac.uk}

\affiliation{Department of Chemistry, University of Cambridge, Cambridge CB2 1EW, United Kingdom}
\date{\today}

\begin{abstract}
        We present a plane-wave basis set implementation of charge constrained 
        density functional molecular dynamics (CDFT-MD) for simulation of 
        electron transfer reactions in condensed phase systems. Following 
        earlier work of Wu {\it et al.} Phys. Rev. A {\bf 72}, 024502 (2005),
        the density functional is minimized 
        under the constraint that the charge difference between donor and 
        acceptor is equal to a given value. The classical ion dynamics is
        propagated on the Born-Oppenheimer surface of the charge constrained state. 
        We investigate the dependence of the constrained energy and of the energy gap
        on the definition of the charge, and present expressions for the constraint 
        forces. The method is applied 
        to the Ru$^{2+}$-Ru$^{3+}$ electron self-exchange reaction in 
        aqueous solution. Sampling the vertical energy gap along CDFT-MD 
        trajectories, and correcting for finite size effects, a reorganization 
        free energy of 1.6~eV is obtained. This is 0.1-0.2 eV lower than 
        a previous estimate based on a continuum model for solvation. 
        The smaller value for reorganization free energy can be explained by 
        the fact that the Ru-O distances of the divalent and trivalent 
        Ru-hexahydrates are predicted to be more similar in the electron transfer 
        complex than for the separated aqua-ions.  
\end{abstract}

\maketitle

\section{Introduction}
        Atomistic simulation of electron transfer 
        reactions has become a major task in computational physics and chemistry. This field 
        was pioneered in the Eighties  
	by the works of Warshel\cite{Warshel82,Hwang87,King90}, 
        Kuharski {\it et al}\cite{Kuharski88,Bader90} and others\cite{Kakitani85_1,Tachiya89,Carter89jpc}. 
        Electron donor and acceptor
	and the solvent were modelled with classical potential energy functions 
        and the diabatic free energy profiles were sampled with (biased) molecular 
        dynamics simulation. 
	These early simulations gave unprecedented insight into electron transfer reactions
	and provided a first numerical confirmation for the validity of the linear response 
	approximation of the Fe$^{2+}$-Fe$^{3+}$ electron self-exchange reaction\cite{Kuharski88}, 
        thus confirming a crucial assumption in Marcus theory of 
        electron transfer\cite{Marcus56_1}.     

         The application of modern density functional molecular dynamics to electron transfer 
         reactions has not been successful until recently\cite{Sit06}. The reason is that common
         exchange-correlation functionals are of limited use for this task due to their
         tendency to erroneously delocalize electrons, thus prohibiting an accurate 
         modelling of charge transfer states.  This deficiency termed electron self-interaction 
         or delocalization error\cite{Zhang98,Mori-Sanchez06_2} is intrinsic to GGA and 
         hybrid density functionals unless special care is taken in their 
         parametrization. In parallel to the development of functionals with minimal  
         self-interaction error\cite{Mori-Sanchez06_1,Cohen07jcp1}, 
         a number of correction schemes have been 
         proposed to minimize the delocalization error of existing and computationally 
         inexpensive density 
         functionals\cite{Perdew81,Tavernelli07,Harrison87,d'Avezac05,VandeVondele05pccp}, 
         including DFT+U\cite{Liechtenstein95,Migliore09,Deskins09}, formulation of a penalty density 
         functional\cite{Sit06}, and constrained 
         DFT\cite{Dederichs84,Akai86,Wu05,Wu06jcp,Wu06jpca,Wu06jctc,Wu07,Wu09,
         Behler05,Behler07,Schmidt08}. 
         Development of such schemes 
         are particularly important in density functional molecular dynamics, where 
         efficient computation of the exchange-correlation energy and forces is 
         absolutely crucial.         
         
         In recent years several groups\cite{Wu09,Schmidt08,Behler07} have greatly 
         advanced the development of the constrained DFT approach. In this method an 
         external potential is added
         to the Kohn-Sham equations preventing an excess electron or hole from 
         wrongly delocalizing over donor and acceptor ions. The external potential is 
         varied until a given constraint on the density, for instance a charge or spin constraint, 
         is satisfied. The search for this external potential, usually carried out within 
         a Lagrange multiplier scheme, introduces a second iteration loop in addition to 
         the usual self-consistent iteration of the Kohn-Sham equations. Due to the 
         localized (or diabatic) nature of the constrained states the delocalization error 
         of common density functionals is reduced, though not eliminated. Hence 
         one can expect that common density functionals describe constrained states
         as well as any other states where delocalizing electrons are not present. 
         
         The construction of charge localized states seems to be a somewhat 
         artificial procedure since the external potential and charge are all but uniquely defined. 
         For large donor-acceptor distances one can expect, however, that the details of charge
         definition are less relevant. Thus it is appealing to use CDFT to describe the charge localized 
         (diabatic) states of long-range electron transfer reactions. However, as it is 
         possible to extract approximate electronic coupling matrix elements between 
         CDFT states\cite{Wu06jcp}, one can construct approximate Hamiltonian 
         matrices in the space of constrained states (CDFT configuration interaction)\cite{Wu07}. 
         The adiabatic states diagonalizing this 
         Hamiltonian suffer less from the delocalization error than unconstrained 
         states and can be used to describe short range phenomena such as chemical 
         bond break reactions\cite{Wu07,Wu09}.     
                         
         All of the above mentioned calculations 
         (except the ones described in Refs.~\cite{Behler05,Behler07})
         were carried out in the gas-phase using quantum chemistry codes. In this work we 
         present a plane-wave basis set implementation 
         of charge constrained density functional molecular dynamics (CDFT-MD) 
         in the Car-Parrinello molecular dynamics code (CPMD)\cite{cpmd}.  This allows us to 
         study the electron self-exchange of the Ru$^{2+}$-Ru$^{3+}$ ion pair in aqueous solution at 
         finite temperature with the solute and solvent treated at the same density functional 
         level of theory. To our best knowledge only one such simulation has been 
         reported previously by Sit {\it et al.}\cite{Sit06}, who developed a penalty 
         density functional approach to simulate the isoelectronic Fe$^{2+}$-Fe$^{3+}$
         electron self-exchange reaction in aqueous solution. A previous attempt to use a 
         charge restraint rather than a charge constraint to drive a charge transfer
         reactions with Car-Parrinello molecular dynamics was also reported\cite{Sulpizi05}. 
         However, it turned out that at reasonably high restraining forces the 
         charge restraint does not provide enough bias for obtaining charge localized states.          
         
         In the present work we define the constraint as the charge difference between the 
         electron donating Ru$^{2+}$-hexahydrate and the electron accepting
         Ru$^{3+}$-hexahydrate. The charge constrained state is energy optimized
         at each molecular dynamics step and the dynamics of the system propagated 
         on the constrained Born-Oppenheimer surface. We then construct 
         Marcus-type free energy profiles by calculation of the vertical energy gap between 
         the two charge transfer states along the molecular dynamics trajectory. The 
         reorganization free energy obtained, about 1.6 eV after correction for finite 
         size effects, is in fair agreement with experimental estimates and other 
         computational studies indicating that CDFT-MD can be successfully applied
         to electron transfer problems in the condensed phase.          
	
	This paper is organized as follows: First we briefly review the theoretical background  
	of CDFT. Then we address the dependency of CDFT energies on the choice of the 
	constraining potential for gas phase and condensed phase electron transfer systems. 
	Thereafter our molecular dynamics implementation of CDFT is validated by 
	finite temperature simulation of the H$_2^+$ molecule. The main objective of 
	the present work, the charge constrained density functional MD simulation of the aqueous 
	Ru$^{2+}$-Ru$^{3+}$ electron self exchange reaction is presented thereafter. 
	The results are discussed in light of experimental results, and previous classical molecular 
	dynamics and continuum studies. At the end of this work we give a 
	perspective on future work planned. In the appendix we 
	give explicit expressions for constraint forces due to the charge constraint
	and summarize relevant technical details of CDFT-MD calculations.

\section{Theory}
\subsection{Charge constrained density functional molecular dynamics}
CDFT and its working equations have been presented previously in a number
of papers\cite{Wu05,Wu06jcp,Wu06jpca,Wu06jctc,Wu07,Wu09,
         Behler05,Behler07,Schmidt08}. Here
we follow the work of van Voorhis and co-workers\cite{Wu05} and give a 
short summary of the equations pertinent to our molecular dynamics 
implementation.   

In CDFT the usual energy functional $E[\rho]$ is minimized with respect to the 
electron density $\rho$ under the condition that the scleronomic constraint
		\begin{equation}
			\int w(\mathbf{r}) \rho(\mathbf{r})\; d\mathbf{r} = N_\text{c},
			\label{eq::cons}
		\end{equation}
is satisfied, where $N_{\text{c}}$ is a real number. The weight function 
$w(\mathbf{r})$ on the left hand side of  Eq.~\ref{eq::cons} defines the constraint,
for instance the charge of an atom, molecule or molecular fragment, or 
the charge difference between groups of atoms. $N_{\text{c}}$ defines the value 
of the constraint. Both quantities are input parameter that remain unchanged during 
the constrained minimization. Using the Lagrange multiplier technique the new 
energy functional to be minimized is given by
		\begin{equation}
			\label{eq::Wfunc}
			W[\rho,V]=E[\rho]+V (\int w(\mathbf{r}) \rho(\mathbf{r})\; d\mathbf{r}-N_\text{c}),
		\end{equation}
where $V$ is an as yet undetermined Lagrange multiplier. In order to find the minimum of 
$W[\rho,V]$ with respect to (wrt) $\rho$ {\it and} $V$, $W$ is minimized with respect to $\rho$ for a given $V$, 
the gradient 
\begin{equation}
\frac{\delta W}{\delta V}\Big|_\rho =  \int w(\mathbf{r}) \rho(\mathbf{r})\; d\mathbf{r} - N_{\text{c}}
\label{eq::dwdv}
\end{equation}
at the minimizing density is determined to generate the next iteration step for $V$, 
$W$ is minimized for the new value of $V$ wrt $\rho$  and 
this procedure is repeated self-consistently until convergence is reached, that is when
$\delta W/\delta V|_\rho\!=\!0$ and $\delta W/\delta \rho |_V\!=\!0$. 
Efficient Newton methods can be employed for this minimization procedure since second 
derivatives of $W$ wrt $V$ are available  
analytically via density functional perturbation theory\cite{Wu05} or numerically 
as finite differences of Eq.~\ref{eq::dwdv}. For practical applications we define 
in addition to the usual wavefunction convergence criterion a second convergence parameter
for the constraint.
		\begin{equation}
			\label{eq::cccrit}
			\mathcal{C}\geq \left\vert \int w(\mathbf{r}) \rho(\mathbf{r})\; d\mathbf{r}-N_\text{c}\right\vert .
		\end{equation}
$\mathcal{C}$ is a measure of how accurately the charge constraint Eq.~\ref{eq::cons} 
is fulfilled.

It is interesting to note that Eq.~(\ref{eq::Wfunc}) implies the following 
exact relation for the energy difference between two constrained states 
A and B,  $\Delta E_{\text{AB}}$, and the continuous set of Lagrange multipliers connecting 
the two states, $V(N_{\text{c}})$.  
\begin{equation}
\label{eq::consist}
\Delta E_{\text{AB}}=-\int_\text{A}^\text{B} V(N_{\text{c}}) \; dN_\text{c},
\end{equation}
The Lagrange multiplier can thus be interpreted as the force along the charge
coordinate $N_{\text{c}}$. The parallel to thermodynamic integration is intriguing.
		
CDFT-MD simulation or constrained geometry optimization can be carried out provided one takes into 
account the additional forces $\mathbf{F}_{\text{c}i}$ arising from the 
constraint term on the right hand side of Eq.~\ref{eq::Wfunc}. 
Adopting the Hellmann-Feynman theorem the total force on atom 
$i$ at position $\mathbf{R}_i$ is given by 
\begin{equation}
\mathbf{F}_{\text{tot},i} = \mathbf{F}_i +  \mathbf{F}_{\text{c}i} 
\end{equation}
where $\mathbf{F}_{\text{tot},i}\!=\!-\partial W/\partial \mathbf{R}_i$, 
$\mathbf{F}_i\!=\!-\partial E/\partial \mathbf{R}_i$ and 
\begin{equation}
\mathbf{F}_{\text{c}i}= - V \int \rho(\mathbf{r})
\frac{\partial w(\mathbf{r},\mathbf{R})}{\partial \mathbf{R}_i}\;d\mathbf{r}
\label{eq::force}
\end{equation}
Note that the weight function can depend on the coordinates $\mathbf{R}$ 
of all atoms in the system. An explicit expression for the derivative in 
the integrand of Eq.~\ref{eq::force} using the 
weight function defined below is given in appendix~\ref{app::cforce}. In molecular dynamics
simulation the Lagrange multiplier $V$, the total energy and forces have to be
calculated at each time step. The computational  bottleneck is the 
iterative search for $V$. To improve the efficiency we have 
implemented an extrapolation scheme for the Lagrange multiplier 
using Lagrange polynomials. For details we refer to appendix \ref{app::vpred}.

\subsection{Definition of the charge constraint}
The charge constraint is fully defined by the weight function $w$ in the 
integrand of Eq.~\ref{eq::cons} and the actual value of the constraint, $N_{\text{c}}$. 
In principle there are an infinite number of ways of how to choose the weight function. 
In practice one chooses the weight so that Eq.~\ref{eq::cons} corresponds to
some common charge definition. One is then left to investigate how much the 
results depend on the charge definition used. In previous work Mulliken, Loewdin 
and a Becke real space integration scheme have been tested\cite{Wu06jctc}. 
While for short donor-acceptor distances the energy of the constrained state was 
strongly dependent on the weight used, for medium to large distances the dependence
on the weight was reasonably small. The real space density integration scheme was
found to give best overall performance\cite{Wu06jcp,Wu07}.  

Building on this previous work we use a slightly different real space density integration 
scheme for charges, the one according to Hirshfeld\cite{Hirshfeld77}. 
The Hirshfeld charge $q_i$ of an atom $i$ at position $\mathbf{R}_i$ 
is obtained by integration of the total electron 
density $\rho$ multiplied with an atom centered weight function $w_i$,  		
\begin{eqnarray}
q_i & = & Z_i - \int w_i(\mathbf{r},\mathbf{R}) \rho(\mathbf{r})\; d\mathbf{r}  \label{eq::qreal}\\
w_i (\mathbf{r},\mathbf{R}) & =  & \frac{\rho_i (\mathbf{r}-\mathbf{R}_i)}{\sum_{j=1}^{N} \rho_j(\mathbf{r}-\mathbf{R}_j)}
\label{eq::wi}
\end{eqnarray}
where $Z_i$ is the core charge in pseudopotential calculations (or the charge of the nucleus in all-electron 
calculations), and $N$ is the total number of atoms.
The weight function is constructed from the unperturbed promolecular densities of 
atoms $i$, $\rho_i$, 
\begin{equation}
\rho_i (\mathbf{r}-\mathbf{R}_i) = \rho_i(r) 
=\sum_j n_j \vert \psi_i^j(r)\vert^2 \label{eq::rhoi}
\end{equation}
where $r\!=\!|\mathbf{r}-\mathbf{R}_i|$. 
The sum ranges over the radial part of the promolecular atomic orbitals 
$\psi_i^j(r)$ with occupation number $n_j$. The weight $w_i$ is close to unity up 
to a distance of about one atomic 
radius and goes to zero according to the decay of the promolecular atomic density.
Note that the sum of the Hirshfeld charges of all atoms is equal to the total charge of the 
system.  

\begin{figure}[ht]
\centering
\includegraphics[clip=true,scale=0.25]{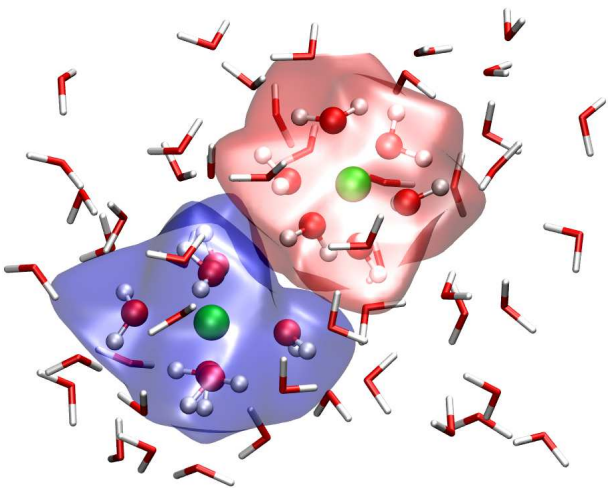}
\caption{
Isosurfaces of the weight function Eq.~\ref{eq::w} for values $w\!=\!0.6$ (red) and 
$w\!=\!-0.6$ (blue). The snapshot is taken from a charge constrained MD simulation of the 
Ru$^{2+}$-Ru$^{3+}$ electron transfer complex in aqueous solution. Ru ions are depicted 
as green spheres and first shell water molecules are depicted in ball and stick 
representation. All atoms composing the Ru-hexahydrate in the upper right (lower left) part of 
the figure are donor (acceptor) atoms. Solvent molecules not included in the charge
constraint are depicted in stick representation.}
\label{fig::weight}
\end{figure}

A natural choice for the constraint for electron transfer reactions is the charge
difference between a set D of atoms comprising the electron donor and a set A
of atoms comprising the electron acceptor. 
\begin{equation}
C - \int w(\mathbf{r}) \rho(\mathbf{r})\; d\mathbf{r} = 
\sum_{i\in D} q_i - \sum_{i\in A} q_i = N_{\text{c}} \label{eq::cons1}
\end{equation}
where  
\begin{equation}
w=w_\text{D}-w_\text{A}=\frac{\sum_{i \in \text{D}} \rho_i(\mathbf{r}-\mathbf{R}_i)-\sum_{i \in \text{A}} \rho_i(\mathbf{r}-\mathbf{R}_i)}{\sum_{j=1}^{N} \rho_j(\mathbf{r}-\mathbf{R}_j)}\label{eq::w}.
\end{equation}
and $\rho_i$ is given by Eq.~\ref{eq::rhoi}. The constant $C$ in Eq.~\ref{eq::cons1} is equal to the 
difference in the core charges, $C\!=\!\sum_{i\in D} Z_i - \sum_{i\in A} Z_i$, merely causing 
a  constant shift of $W$ (Eq.~\ref{eq::Wfunc}) by $VC$. 
The weight function Eq.~\ref{eq::w} used for the simulation of the Ru$^{2+}$-Ru$^{3+}$ 
electron self-exchange reaction in aqueous solution is illustrated in Fig.~\ref{fig::weight}. 
The donor (acceptor) atoms are comprised of Ru$^{2+}$ (Ru$^{3+}$) and all atoms of the
first solvation shell of Ru$^{2+}$ (Ru$^{3+}$) . Note that the sign of the weight function 
changes sharply at the interface of the donor-acceptor complex.   
Assuming that a charge equivalent to one electron is transferred, the constraint 
value $N_{\text{c}}$ is set equal to 1 for the initial state and equal to -1 for the final state. 

While relying on the basic charge definition Eqs.~\ref{eq::qreal}-\ref{eq::wi}, we have 
investigated different functional forms for the densities $\rho_i$ that define the atomic 
weight function $w_i$ of Eq.~\ref{eq::wi}. The results are presented in 
section~\ref{sec::wdep}. Technical details concerning the calculation of the weight 
function can be found in appendix \ref{app::wco}.
		
\subsection{Electron transfer free energy curves\label{subsec:ET}}
In Marcus theory\cite{Marcus56_1}
electron transfer reactions are described by two charge localized or diabatic 
free energy curves, one for the reactant state (A) and one for the 
product state (B). Electron transfer is assumed to occur at  
the crossing point of the two curves. In the limit of linear response
the two free energy curves are parabolic and the activation free energy
for electron transfer is given by
\begin{equation}
\Delta A^\ddagger = \frac{(\Delta A + \lambda)^2}{4\lambda}
\end{equation}   
where $\Delta A$ is the reaction free energy and $\lambda$ is the reorganization
free energy. The latter is inversely proportional to the curvature of the free energy
parabola and is a measure for the free energy required to distort the equilibrium 
configuration of one diabatic state to the equilibrium configuration of the 
other diabatic state while staying on the same free energy curve.   
For electron self-exchange reactions $\Delta A\!=\!0$ and the
activation free energy is entirely determined by the reorganization 
free energy.  

Following Warshel\cite{Warshel82} 
the diabatic free energy curves defining the reorganization free energy 
can be obtained by sampling the vertical energy gap $\Delta E$ 
\begin{equation}
\Delta E(\mathbf{R})=E_\text{B}(\mathbf{R})-E_\text{A}(\mathbf{R})\label{eq::de}
\end{equation}
using molecular dynamics simulation. $E_\text{A}(\mathbf{R})$ and $E_\text{B}(\mathbf{R})$ 
are the charge localized (diabatic) potential energy surfaces and the vector 
$\mathbf{R}$ denotes the coordinates of all atoms in the system. The relative Landau free energy 
$A_M(\Delta E)$ along this coordinate is given for state $M$, $M$\,=\,A, B, by 
\begin{equation}
\label{eq::freeEn}
A_M(\Delta E)=-k_\text{B}T \ln p_M(\Delta E),
\end{equation}
where $k_\text{B}$ is the Boltzmann constant, $T$ the temperature and $p_M(\Delta E)$ 
the probability distribution of the reaction coordinate in state $M$. If the free energy
curves are parabolic the reorganization free energy is equal to the average vertical energy
gap,  
\begin{equation}
\lambda = \langle \Delta E \rangle_{\text{A}}
\label{eq::lambda}
\end{equation}
where $\langle \cdots \rangle_{\text{A}}$ denotes the usual canonical average 
for state A.  
 		
The free energy curves Eq.~\ref{eq::freeEn} can be obtained by sampling configurations
with charge constrained density functional molecular dynamics in state A as 
described above, followed by calculation of the vertical energy gap Eq.~\ref{eq::de} 
between the constrained states A ($N_{\text{c}}\!=\!1$, Eq.~\ref{eq::cons1}) and 
B ($N_{\text{c}}\!=\!-1$) for the set of sampled configurations.   
However, unbiased equilibrium simulations give only accurate results close to the free 
energy minimum of $A_M$ and are of limited use for regions of $\Delta E$ far away 
from the minimum. Fortunately, due to the exact  linear free energy relation between 
the free energy gap and the energy gap\cite{Warshel82,Tachiya89,Tateyama05,Blumberger06jcp}, 
\begin{equation}
A_\text{B}(\Delta E)-A_\text{A}(\Delta E) = \Delta E
\label{eq::lin}
\end{equation}
it is possible to calculate a good part of the curve at high free energies accurately from 
equilibrium simulations. Thus, using information from two distinct regions of 
$\Delta E$ one can construct a reasonably accurate free energy profile without 
the use of computationally expensive enhanced sampling 
methods\cite{Blumberger06jcp,Blumberger08jacs}.  

\section{Computational details\label{sec::comp}}
Charge constrained density functional molecular dynamics has been implemented
in the CPMD code\cite{cpmd}. Unless stated otherwise, all calculations 
were carried out  with the BLYP\cite{Becke88,Lee88} functional using a reciprocal space 
cutoff of $70$ Ry, Troullier-Martins pseudopotentials for nuclei+core 
electrons\cite{Troullier91}, pseudoatomic densities for construction of the weight 
function Eq.~\ref{eq::w}, a convergence criterion for the wavefunction gradient 
of $1\times 10^{-5}$H and a charge constraint convergence criterion  defined in 
Eq.~(\ref{eq::cccrit}) of $\mathcal{C}=5\times 10^{-5}$e. All calculations were 
carried out in the lowest spin state. 

Ru$^{2+}$-Ru$^{3+}$ electron self-exchange was simulated in a 
periodic box of dimension $14.5 \times 11.35 \times 11.35$\AA~ containing two 
Ru ions and 63 water molecules.  The donor and acceptor groups are 
comprised of the ion and the six water molecules forming the first 
solvation shell, Ru(H$_2$O)$_6^{2+}$ and Ru(H$_2$O)$_6^{3+}$, respectively,
see Figure~\ref{fig::weight}. The charge constraint Eq.~\ref{eq::cons1} was 
set equal to 1 corresponding to the reactant diabatic state A. 
Pseudoatomic densities for construction of the weight 
function Eq.~\ref{eq::w} were used, i.e. the charges correspond to the definition
of Hirshfeld\cite{Hirshfeld77}.  
The Ru$^{2+}$ and Ru$^{3+}$ aqua ions are both low spin
as opposed to the corresponding aqua-ions of Fe$^{2+}$ and Fe$^{3+}$\cite{Sit06}.  
Hence, the dublet state was chosen for simulation of the aqueous ET complex. 
The initial configuration was taken from an equilibrated 
classical molecular dynamics trajectory carried out in a previous 
investigation of the same reaction\cite{Blumberger08mp}. The distance between 
the two Ru ions was fixed at 5.5~\AA~ using the RATTLE algorithm \cite{Andersen83}.  
The system was simulated in the NVT ensemble at $300$\,K using a 
chain of Nose-Hoover thermostats\cite{Martyna92} of length $4$ with 
a frequency $1000$\,cm$^{-1}$. 
To increase the efficiency of charge constrained Born-Oppenheimer molecular 
dynamics we used an extrapolation scheme for prediction of the Lagrange 
multiplier $V$ as described in appendix~\ref{app::vpred}, an extrapolation scheme 
for the initial guess of the wavefunction, slightly higher convergence criteria 
of $\mathcal{C}=5\times 10^{-4}$\,e for the constraint convergence 
and of $2\times 10^{-5}$\,H for the wavefunction gradient, and an MD time step 
of  $40\,\text{au}\!=\!0.96$\,fs. With this setup the average number of iterations 
for $V$ were  $\sim 2\!-\!3$ per molecular dynamics 
time step. Accordingly, the computational overhead compared to standard Born-Oppenheimer
molecular dynamics without charge constraint is about a factor of 2-3.
The average drift of the conserved energy was $-9.7\times10^{-5}$H/atom/ps along
a trajectory of length 6.6 ps.  This is somewhat large  but still acceptable 
for our purposes. Better energy conservation can be obtained if tighter 
convergence criteria and a smaller time step is used.  For calculation
of thermal averages the first ps of dynamics was discarded. The energy 
gap Eq.~\ref{eq::de} was calculated for 225 equidistantly spaced
snapshots taken from the last 5.5 ps of the trajectory using the same 
convergence criteria as for the MD simulation. For calculation of the 
product diabatic state B the constraint Eq.~\ref{eq::cons1} was set equal to -1.

\section{Results and discussion}
Before we present the results for electron self-exchange in the condensed 
phase we report on a series of test calculations carried out for simple
electron transfer systems in the gas phase. These include a basic validation
of charge constrained single point calculations and molecular dynamics, 
calculation of the dissociation curve of H$_2^+$ and He$_2^+$ to test
the correct long-range behaviour of CDFT, and an investigation of the dependence 
of the constrained state energies on the weight function used.   

\subsection{(CO)$^-$}
\begin{figure}[ht]
\centering
\includegraphics[scale=0.3,clip=true]{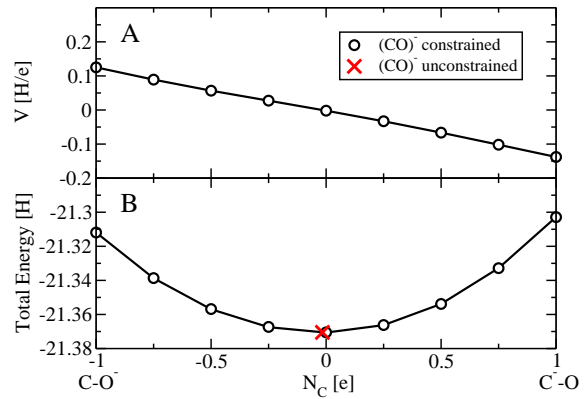}
\caption{Lagrange multiplier $V$ (A) and potential energy $E$ (B) versus 
constrained charge difference $N_\text{c}$ between C and O atom in (CO)$^-$. 
In (B) the computed 
energies are indicated by circles and the 
running integral $-\int_{-1}^{N_\text{c}} V\; dN_\text{c}+E[\text{C}^-\text{O}]$ by 
straight lines. The potential energy and charge difference for unconstrained 
wavefunction optimization of (CO)$^-$ is indicated by a cross.}
\label{fig::co-}
\end{figure}

As a first test of our implementation we carried out a series of   
charge constrained wavefunction optimizations on the (CO)$^-$ molecule. 
The distance between the C and the O atom was chosen to be $2$\AA{} in 
order to avoid the problem of spin contamination which occurs 
for larger distances. The oxygen atom is chosen as the electron donor 
and the carbon atom as the electron acceptor. The excess electron is transferred 
from the oxygen to the carbon atom by increasing the 
charge difference $N_\text{c}$ (Eq.~\ref{eq::cons1}) between the atoms 
in small steps from -1 to 1. The Lagrange multiplier $V$ and the potential energy $E$ 
of the constrained states are shown in Fig.~\ref{fig::co-} as a function of $N_{\text{c}}$.

According to Eq.~\ref{eq::consist} the energy difference between C$^-$O 
and CO$^-$ should be equal to the negative of the integral of $V$ over $N_\text{c}$. 
Indeed, the difference between the end-points is 
$\Delta E_{\text{AB}} = E[\text{C}^-\text{O}]-E[\text{CO}^-]\!=\! 9.038$ mH whereas 
integration gives $-\int_{-1}^{1} V\; dN_\text{c}\!=\!9.084$ mH. The numbers match 
within the given convergence criteria. Furthermore, the potential energy obtained from 
unconstrained calculations (denoted by a cross in Fig.~\ref{fig::co-}) lies on 
the constrained potential energy curve at $N_{\text{c}} \approx 0$.  
Thus, the unconstrained state can be reproduced by constrained 
wavefunction optimization if a constraint value is used that is equal to the charge
difference of the unconstrained state.         

\subsection{Dissociation of H$_2^+$ and He$_2^+$}
The wrong dissociation curve for H$_2^+$ is probably one of the most spectacular
failures of GGA and hybrid density functionals. The reason for this failure 
is well known\cite{Zhang98}. Due to the wrong scaling behaviour of these functionals wrt
electron number the charge delocalized state is predicted to be lower in energy
than the charge localized state, even though the two states are degenerate 
in the limit of large inter-nuclear separation distance in exact theory. Using 
CDFT configuration interaction Wu {\it et al.} could circumvent this 
problem\cite{Wu07}. The authors reported 
a dissociation curve that matched the exact Hartree-Fock curve remarkably
well at the equilibrium region and at long range. 

Here we calculate the dissociation curve for 
H$_2^+$ and He$_2^+$ using a single charge constrained state in order to 
test the correct long range behaviour of our CDFT implementation. The constraint
is again the charge difference between the two atoms. In case of He$_2^+$  $N_\text{c}$ 
was set equal to 1 for all distances. For H$_2^+$ the constraint value
was set equal to the charge difference obtained when the 
promolecular pseudoatomic reference orbitals were used for construction 
of the initial wavefunction\cite{Wu07}. This was necessary because a value 
$N_\text{c}\!=\!1$ would only be obtained at very high external potentials 
for which the wavefunction would not converge. Thus, the constraint value 
for  H$_2^+$ changes with distance from about $0.86$ at $2$\AA{} to 1.0 at 
$4.4$\AA{} and larger distances.
			
\begin{figure}[ht]
	\centering
	\includegraphics[scale=0.4,clip=true]{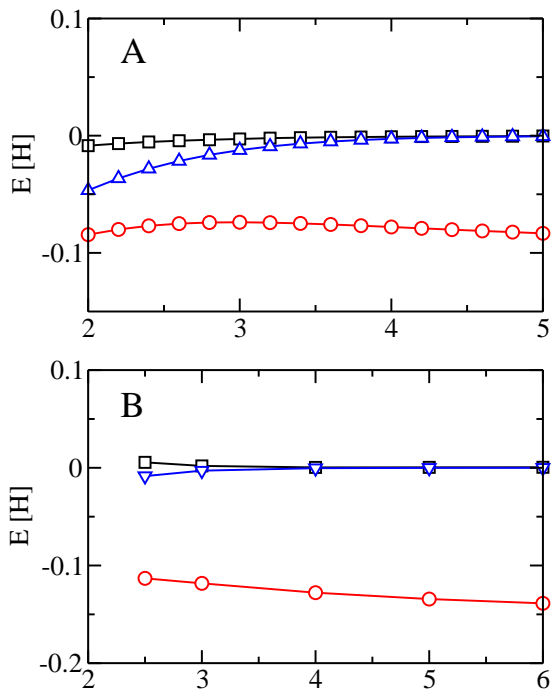}
\caption{Dissociation curves for H$_2^+$ (A) and He$_2^+$ (B). 
	Constrained and unconstrained DFT results are indicated by squares and circles, respectively.
	The exact Hartree-Fock curve for H$_2^+$ and the curve for He$_2^+$ at the FCI/aug-cc-pV5Z level of 
	theory\cite{Pieniazek07} are indicated by triangles, respectively. }
\label{fig::h2+-Zn2+}
\end{figure}

The energies of the charge constrained states of H$_2^+$ and He$_2^+$  
relative to the (unconstrained) energy of the isolated fragments are 
shown in Fig.~\ref{fig::h2+-Zn2+}, together with the exact Hartree Fock 
curve for H$_2^+$ and the essentially exact FCI curve for He$_2^+$\cite{Pieniazek07}.
The CDFT curves agree well with the exact curves for distances larger than ~3\AA. This 
is a considerable improvement with respect to the results of unconstrained 
optimizations, which give the wrong dissociation limit. It also 
shows that the optimization in the external potential yields the correct electronic 
state of the unconstrained isolated fragments.
The significant deviation for smaller distances is due to the fact that 
only a single charge localized state is used, which can not describe 
chemical bonding between the fragments. 
The deficiency at short range can be cured using the CDFT-configuration 
interaction approach introduced in Ref.~\cite{Wu07}.      
 
\subsection{Dependence on the weight function}
		\label{sec::wdep}
A crucial issue in CDFT calculations is the dependence of the numerical results 
on the definition of the charge constraint. While relying on 
the basic real-space charge definition Eqs.~\ref{eq::qreal}-\ref{eq::wi}, we have 
investigated different functional forms for the densities $\rho_i$ that define the atomic 
weight function $w_i$ of Eq.~\ref{eq::wi}. Besides numerical 
pseudoatomic densities we have tested a minimal basis of Slater functions and 
a minimal basis of Gaussian functions,  
\begin{equation}
\label{eq::wgauss}
\rho_i(\mathbf{r}-\mathbf{R}_i)=\frac{N_{\text{el},i}}{\sigma_k 
\sqrt{2 \pi}}e^{-\frac{(\mathbf{r}-\mathbf{R}_i)^2}{2 \sigma_k^2}},
\end{equation}
where $\sigma_k$ is the width of particle species $k$ and $N_{\text{el},i}$ 
denotes the number of electrons of the isolated atom $i$.  
Using Gaussian functions one can vary the decay of the weight functions 
systematically simply by changing the exponent of the densities in Eq.~\ref{eq::wgauss}. 
For slowly decaying densities the weight function changes sign smoothly, while 
for densities that match the pseudoatomic densities the sign changes sharply. 
In the limit of an infinitely fast decaying density $\rho_i$ the electron 
density $\rho$ at a given point in space is assigned to the atom closest 
to this point. This corresponds to the charge definition according to Voronoi.
For the real space integration of Eq.~\ref{eq::qreal} we have introduced a 
radial cutoff $R_\text{c}$ (see appendix~\ref{app::wco}). Its influence 
on the constrained energy is also reported here. 
\begin{table}
\caption{Charge constrained energies for hole transfer in Zn$_2^+$. Pseudoatomic, 
Slater and Gaussian denote the functional form of the densities $\rho_i$ (Eq.~\ref{eq::rhoi}) 
that define the atomic weight function $w_i$ of Eq.~\ref{eq::wi}. $\sigma$ is the width of 
the Gaussian function Eq.~\ref{eq::wgauss} and $R_\text{c}$ denotes the radial weight cutoff
(see appendix~\ref{app::wco}).}
\label{table::zn}
		\begin{center}
		\begin{tabular}{|c|c|c|c|}\hline
		\multicolumn{4}{|c|}{$r=4.0$\AA{}}\\\hline
			weight & $\sigma$ [\AA{}] & $R_\text{c}$ [\AA{}] & Energy [H] \\ \hline\hline
			\multirow{2}{*}{pseudoatomic} &  & $3.762$ & $-113.7834$\\
				&  & $5.121$ & $-113.7834$\\ \hline
			\multirow{2}{*}{Slater} & & $3.762$ & $-113.7780$\\
				& & $5.121$ & $-113.7783$\\ \hline
			\multirow{7}{*}{Gaussian}& $0.5$ & $3.762$ & $-113.7859$\\ 
				& $0.732$ & $3.762$ & $-113.7849$\\
				& $1.0$ & $3.762$ & $-113.7818$\\
				& $2.0$ & $3.762$ & $-113.7633$\\
				& $1.0$ & $2.56$ & $-113.7813$ \\
				& $1.0$ & $3.0$ & $-113.7814$\\
				& $1.0$ & $5.0$ & $-113.7819$\\ \hline
		\end{tabular}
		\end{center}

		\begin{center}
		\begin{tabular}{|c|c|c|c|}\hline
		\multicolumn{4}{|c|}{$r=2.5$\AA{}}\\\hline
			weight & $\sigma$ [\AA{}] & $R_\text{c}$ [\AA{}] & Energy [H] \\ \hline\hline
			\multirow{2}{*}{pseudoatomic} &  & $3.762$ & $-113.7744$\\
			&  & $5.121$ & $-113.7745$\\ \hline
			\multirow{2}{*}{Slater} & & $3.762$ & $-113.7471$\\
				& & $5.121$ & $-113.7473$\\ \hline
			\multirow{7}{*}{Gaussian}& $0.5$ & $3.762$ & $-113.7824$\\ 
				& $0.732$ & $3.762$ & $-113.7755$\\
				& $1.0$ & $3.762$ & $-113.7613$\\
				& $2.0$ & $3.762$ & $-113.6495$\\
				& $1.0$ & $2.56$ & $-113.7781$ \\
				& $1.0$ & $3.0$ & $-113.7612$\\
				& $1.0$ & $5.0$ & $-113.7617$\\ \hline
		\end{tabular}
		\end{center}
\end{table}
The charge constrained energies of Zn$_2^+$ obtained for different weight functions 
are summarized in table~\ref{table::zn}. The Zn-Zn distance was chosen to be $r=4$\,\AA{} 
and $r=2.5$\,\AA. The latter is equal to twice the covalent radius of Zn. 
The constraint was again the charge difference, $N_{\text{c}}\!=\!1$.
Considering hole transfer at a distance $r=4$\,\AA, we find 
that the change in energy is only a few mH when the width of the 
Gaussian function is varied from 0.5 to 1.0\,\AA. If 
larger values for the widths are used the weight function 
becomes unphysically smooth and the energy increases. The energy 
obtained with pseudoatomic functions differ by not more than 2\,mH 
from the energies obtained with Gaussian functions. A somewhat 
larger deviation is obtained for Slater functions. The cutoff 
$R_{\text{c}}\!=\!3.762$\,\AA{} for truncation of the weight function 
is sufficient for all weight functions. Overall, the dependence 
of the results on the weight function used is reasonably small. 
This is not the case for hole transfer at a smaller distance
of $r=2.5$\,\AA. Here the details of the weight function are important.
The variation in energy are a few ten mH, an order of magnitude 
larger than at $r=4$\,\AA. Hence, our results indicate that charge constrained
states are well defined only if the distance between donor and acceptor
is larger than at least the sum of their covalent radii.
	
As we are primarily interested in the aqueous 
Ru$^{2+}$-Ru$^{3+}$ electron self-exchange reaction 
(see section~\ref{sec::ru2+-ru3+}) we have 
investigated the dependence of the vertical energy gap Eq.~\ref{eq::de} 
of this system on the functional form of the weight used. 
For this purpose we have taken a snapshot from the constrained MD 
simulation of aqueous Ru$^{2+}$-Ru$^{3+}$ and calculated the two 
charge constrained states $E_{\text{A}}$ ($N_{\text{c}}\!=\!-1$) 
and $E_{\text{B}}$ ($N_{\text{c}}\!=\!-1$). For the Gaussian weight function 
$\sigma_\text{H}=\sigma_\text{O}=0.6$\AA{} and $\sigma_\text{Ru}=1.0$\AA{} is used. 
See section~\ref{sec::comp} for further details.  
The results are summarized in table~\ref{table::ru}. 
\begin{table}
\caption{Dependence of the energy gap Eq.~\ref{eq::de} on the weight
function for the aqueous 
Ru$^{2+}$-Ru$^{3+}$ electron self-exchange reaction. The charge
is the sum of the atomic charges of Ru$^{3+}$(H$_2$O)$_6$ in the 
final ET state.}
\label{table::ru}
		\begin{center}
		\begin{tabular}{|c|c|c|}\hline
			weight & energy gap [eV] & charge [e] \\ \hline\hline
			pseudoatomic & $1.587$ & $1.5663$ \\ \hline
			Slater & $1.732$ & $1.2639$ \\ \hline
			Gaussian & $1.552$ & $1.7764$ \\ \hline
		\end{tabular}
		\end{center}
\end{table}
The energy gaps differ by less than $0.18$ eV for the three weights considered
and are within $0.04$ eV for pseudoatomic and Gaussian weight functions. This
variation is not insignificant and should be considered as a lower limit
of the error of the results presented in section~\ref{sec::ru2+-ru3+}. 
More than the gap energies varies the charge of the electron donor 
Ru$^{2+}$(H$_2$O)$_6$ (recall that only the charge {\it difference} between donor
and acceptor is constrained). The sensitivity of the results on the 
weight used is probably a consequence of the close approach of the 
first shell water molecules that bridge the two Ru-ions. These water molecules
form strong hydrogen bonds that make the constrained energy and charges 
susceptible to details of the weight function at the interface of the two 
Ru-complexes.
 
\subsection{CDFT-MD for H$_2^+$}
\begin{figure}[hb]
\centering
\includegraphics[scale=0.3,clip=true]{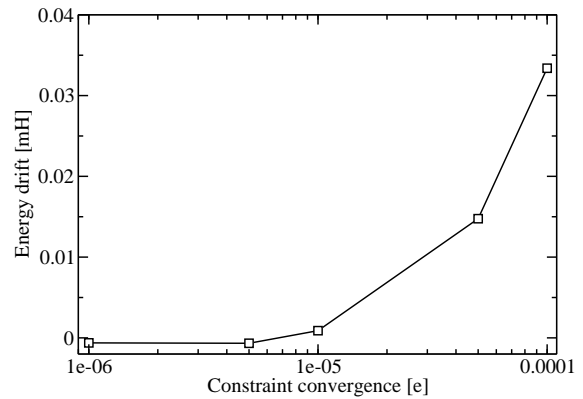}
\caption{CDFT-MD simulation of H$_2^+$. The linear drift in the conserved 
energy is shown as a function of the charge constraint convergence criterion 
$\mathcal{C}$ defined in Eq.~\ref{eq::cccrit}.}
\label{fig::h2+edrift}
\end{figure}

The molecular dynamics implementation of CDFT is tested by simulating an isolated H$_2^+$ molecule 
on the Born-Oppenheimer surface of a single charge-constrained state. A 
charge difference $N_{\text{c}}\!=\!0.5$\,e is enforced giving an average charge of 
$0.25$\,e respectively $0.75$\,e on the two H atoms.
The system is simulated in the NVE ensemble at a temperature of approximately $300$\,K 
using the Velocity Verlet algorithm.  
A value of $5\times 10^{-6}$\,H for the convergence of the wavefunction 
gradient is used and a timestep of $20\text{au}\approx0.48$\,fs. In order to assess 
the dependence of energy conservation on the convergence criterion for the charge 
constraint, Eq.~\ref{eq::cccrit}, we calculated a series of trajectories of length 
1 ps for different values of $\mathcal{C}$. The total 
linear drift of the conserved energy as a function of $\mathcal{C}$ is shown 
in Fig.~\ref{fig::h2+edrift}. 
The drift is less than $1 \times 10^{-3}$ mH/atom/ps at 
$\mathcal{C}\!=\!1\times 10^{-6}$\,e 
showing that the total energy is essentially conserved if a tight convergence criterion 
for the charge constraint is applied. 
The sharp rise of the total energy drift at a value of $\mathcal{C}\!=\!1\times 10^{-5}$\,e 
is due to the fact that the system essentially behaves like a harmonic oscillator. 
Small deviations from the target constraint value can lead to a resonance between the bond 
vibration and the external constraint potential. This can cause instabilities in the 
integration of the equations of motion, which can even lead to dissociation of the molecule. 
Fortunately, this behaviour is rather exceptional. 
The resonance effect is dampened in larger systems allowing us to use less strict 
convergence criteria than for H$_2^+$.

\subsection{Ru$^{2+}$-Ru$^{3+}$ electron self-exchange in aqueous solution}
		\label{sec::ru2+-ru3+}
We finally present our results for the CDFT-MD simulation of Ru$^{2+}$-Ru$^{3+}$ electron
self-exchange in the condensed, aqueous phase. The charge constraint is chosen as the 
charge difference between the electron donating group, Ru$^{2+}$(H$_2$O)$_6$, and 
the electron accepting group, Ru$^{3+}$(H$_2$O)$_6$, and the constraint value is
$N_{\text{c}}\!=\!1$. This choice is motivated by the fact that one wants to 
transfer a charge corresponding to one electron from the donor to the acceptor
complex. The six water molecules forming the first coordination shell are included in 
the constraint as the redox active orbitals are delocalized over the metal and the 
first coordination shell. The distance of the two Ru ions is constrained to 5.5~\AA.
The same distance was used in a previous classical molecular dynamics 
simulation of this reaction\cite{Blumberger08mp}. 
No other constraints on the dynamics are imposed. Details for the CDFT-MD 
simulation are summarized in section~\ref{sec::comp}. 
 
During CDFT-MD the highest occupied majority spin orbital (HOMO) of the ET complex 
is correctly located on the donor complex, and the lowest unoccupied minority spin
orbital (LUMO) is located on the acceptor complex. As expected, the two molecular orbitals 
are composed of a $d$ orbital of the metal $t_{2g}$ manifold and the 
$p$ orbitals of the ligands. Similarly to the separated aqua ions, there is no 
mixing with orbitals of solvent molecules beyond the first solvation shell.    

As the charge difference is constrained, only, the absolute charge of donor and 
acceptor complex is free to vary during the dynamics run. The charge 
fluctuations are very small, however, $\sigma=0.05$\,e. The average 
charge of the Ru$^{2+}$(H$_2$O)$_6$ complex, $0.52$\,e, and of the 
Ru$^{3+}$(H$_2$O)$_6$ complex, 1.52\,e, are significantly smaller than their formal 
charges of +2\,e and +3\,e, respectively. They are, however, similar to the charge 
of a single Ru$^{2+}$(H$_2$O)$_6$ (Ru$^{3+}$(H$_2$O)$_6$) ion
in aqueous solution, 0.75\,e (1.15\,e). Thus, the charge constraint localizes an 
excess charge of $-0.23$\,e (0.37\,e) on the donor (acceptor) complex relative to 
the charge of the isolated aqua ions. The charge of the remaining solvent, 2.96\,e,
is more than half of the total system charge. Although some charge transfer between 
the ET complex and the solvent is expected, the magnitude of this effect seems
rather large. The reason for this is not clear, but one may speculate that the 
BLYP functional tends to delocalize the total system charge, an effect
that might be enhanced when periodic boundary conditions are used for simulation of 
the aqueous phase. Yet, the large magnitude of charge transfer to the solvent 
is not a particular feature of CDFT, because this effect already 
occurs for standard (unconstrained) GGA-DFT calculations on solutions 
containing a single ion.  

\begin{figure}[ht]
\centering
\includegraphics[clip=true,width=8cm]{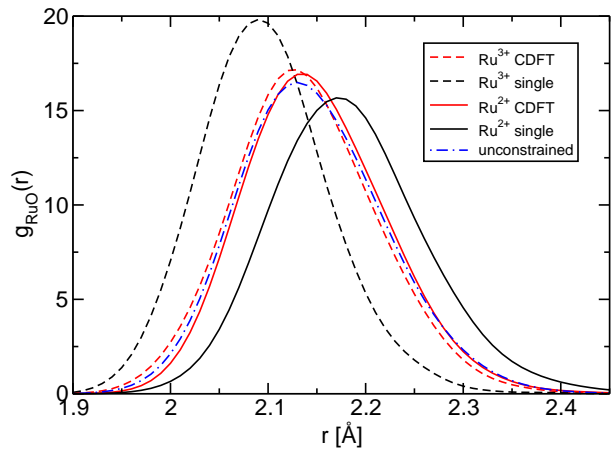}
\caption{First peak of the radial distribution functions of 
Ru$^{2+}$ and Ru$^{3+}$ with the oxygen atoms of the solvent molecules. 
Curves labeled `CDFT' were obtained from  
present CDFT-MD simulations of the solvated ET complex, 
curves labeled `single' were obtained from 
standard Car-Parrinello MD simulation of a single ion in 
aqueous solution\cite{Blumberger06tca,Blumberger05jpcb}, 
and curves labeled `unconstrained' were obtained from 
standard Car-Parrinello MD simulation of the solvated ET complex. 
Note that the curves for Ru$^{2+}$ and Ru$^{3+}$ 
coincide for the latter simulations. All distribution functions were 
smoothed by convolution with a Gaussian of width $0.03$\AA{}.
}
\label{fig::gofrOH-ru23}
\end{figure}

In order to assess the effects of the charge constraint on the coordination geometry, 
we calculated the metal-oxygen radial distribution functions 
($g_\text{RuO}(r)$) of Ru$^{2+}$ and Ru$^{3+}$ in the electron transfer complex 
and compare with the radial distribution function of the single aqueous ions
as obtained from standard (unconstrained) Car-Parrinello molecular dynamics 
simulation\cite{Blumberger06tca,Blumberger05jpcb}. The result is illustrated in 
Fig.~\ref{fig::gofrOH-ru23}. 
In the unconstrained simulations of the single aqua ions the 
average Ru-O bond distances are $2.17$\,\AA{} for Ru$^{2+}$ and 
$2.09$\,\AA{} for Ru$^{3+}$\cite{Blumberger06tca,Blumberger05jpcb}. 
The difference in distance is significantly 
smaller in the ET complex, $2.15$\,\AA{} for 
Ru$^{2+}$ and $2.13$\,\AA{} for Ru$^{3+}$. The deformation of the two complexes
to a more similar coordination geometry must be attributed their strong interactions 
at a rather short Ru-Ru distance of $5.5$\,\AA{}. The solvation shells 
of the two ions interpenetrate and the water molecules bridging the two Ru ions 
form $1\!-\!2$ strong hydrogen bonds during the course of the simulation.
As expected, the solvation structure of the two ions in the ET complex 
is virtually identical if the charge constraint is not imposed. 
The radial distribution functions of the two ions are indistinguishable
(dash dotted lines in Fig.~\ref{fig::gofrOH-ru23}) and the center of the 
peak is located in between the two peaks of Ru$^{2+}$ and Ru$^{3+}$
in the charge constrained ET complex, at $2.14$\,\AA{}. This degeneracy 
is due to the electron delocalization error of the BLYP exchange correlation functional.

\begin{figure}[ht]
\centering
\includegraphics[clip=true,width=8cm]{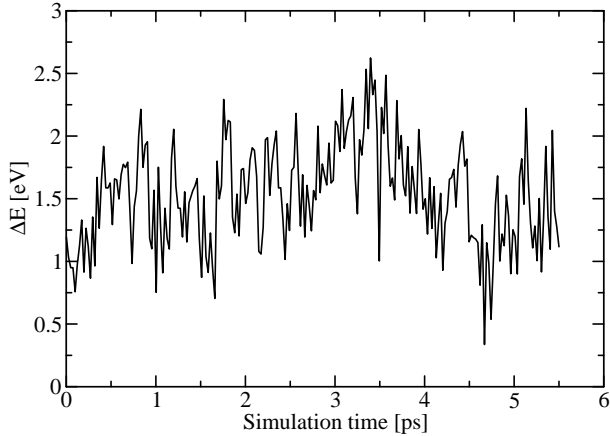}
\caption{Fluctuations of the vertical energy gap $\Delta E$ 
for the Ru$^{2+}$-Ru$^{3+}$ electron self-exchange reaction in aqueous solution.
The energy gap was calculated according to Eq.~\ref{eq::de} for configurations
generated with CDFT-MD. See section~\ref{sec::comp} for details.}
\label{fig::ru23-dE}
\end{figure}

\begin{figure}[hb]
\centering
\includegraphics[scale=0.4,clip=true]{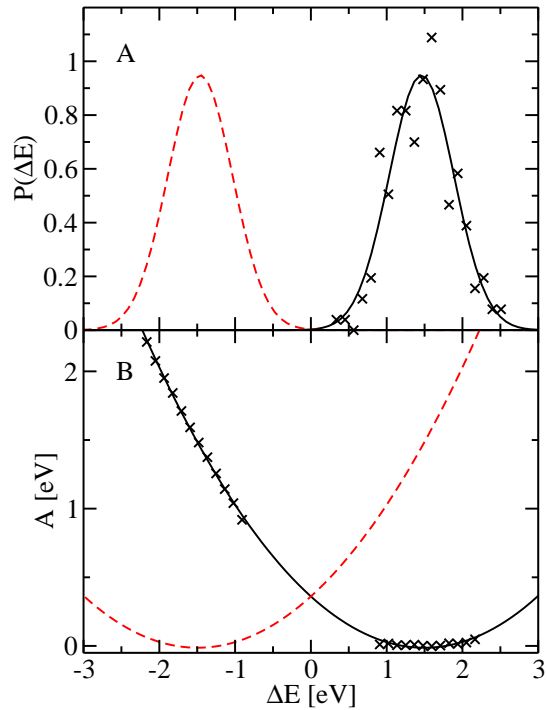}
\caption{Probability distribution of the energy gap Eq.~\ref{eq::de} (A) and diabatic
free energy curves Eq.~\ref{eq::freeEn} (B) for the 
Ru$^{2+}$-Ru$^{3+}$ electron self-exchange in aqueous solution. 
Data points are collected in bins of width $0.11$\,eV. The data points in (A) 
are fit to a Gaussian and reflected about the origin to generate the 
distribution of the symmetrical product state (dashed lines). 
In (B) data points within $1.5$ standard deviations from the center of the 
two distributions are fit to a parabola.}
\label{fig::ru23-fprof}
\end{figure}

The energy gap Eq.~\ref{eq::de} computed for an ensemble of configurations 
taken from the CDFT-MD trajectory, is shown in Fig.~\ref{fig::ru23-dE}
(see section~\ref{sec::comp} for 
computational details). The average is 
$\langle \Delta E\rangle_{\text{A}}\!=\!1.53$~eV and the mean square fluctuation 
is $\langle \delta \Delta E^2\rangle_{\text{A}}^{1/2}\!=\!0.41$~eV. 
The error of the average due to the finite length of the trajectory 
is estimated to be $\approx 0.2$\,eV.
The probability distribution of the energy gap fluctuations 
and the corresponding diabatic free energy profile Eq.~\ref{eq::freeEn}
are shown in Fig.~\ref{fig::ru23-fprof}. Due to the linear free energy 
relation Eq.~\ref{eq::lin} two segments of the free energy curve are obtained 
from the CDFT-MD equilibrium simulation, one for the 
equilibrium region, and one for high free energies at the  
equilibrium region of the product state (see also section~\ref{subsec:ET}
and Refs.~\cite{Blumberger06tca,Blumberger06jcp}). 
The two segments fit well to a parabola with a correlation coefficient of $0.99983$. 
This shows that the Ru$^{2+}$-Ru$^{3+}$ electron-self exchange 
is well described in the linear response approximation, which is 
an essential assumption in Marcus theory of electron transfer. 

Since for electron self-exchange the reorganization free energy equals
the average energy gap (Eq.~\ref{eq::lambda}), 
we obtain $\lambda\!=\!\langle \Delta E\rangle_{\text{A}}=1.53$ eV 
from CDFT-MD. This value includes the reorganization of the two 
Ru-hexahydrates and of 51 water molecules solvating the electron transfer complex.
Reorganization of higher solvation shells and of bulk solvent is missing. 
In previous work we have estimated a correction term for reorganization free energy of the 
missing solvent by carrying out a series of classical molecular dynamics simulations 
for different system sizes and extrapolating the reorganization free energy 
obtained to the limit of infinite dilution\cite{Blumberger08mp}. 
The correction term for the 63 water molecule system amounts to 0.09 eV. 
Our estimate for reorganization
free energy of the infinitely diluted system is then $1.53+0.09\!=\!1.62$~eV.
This estimate is within the range of values obtained in previous classical
molecular dynamics simulations of the same reaction using polarizable 
water models\cite{Blumberger08mp}, 1.60 eV for SWM4-NDP water, 1.71 eV for 
POL3 water and 1.87 eV for AMOEBA water (values taken from 
Ref.~\cite{Blumberger08mp}, corrected for finite size but not for 
nuclear quantum effects). In a continuum study\cite{Rotzinger02} a value of 
1.95\,eV is reported that fits well the experimental rate constant.\cite{Bernhard85}
However, this value was calculated for a Ru-Ru distance of 6.5~\AA~ and should 
be corrected to 1.75 eV if the same distance as in the present CDFT-MD simulations
is assumed (5.5~\AA). Unfortunately a direct experimental estimate for reorganization
free energy is not available since the work term and the electronic coupling 
matrix element are unknown\cite{Bernhard85}. Yet, under a number of assumptions, a value of 
about 2.0 eV was reported in Ref.~\cite{Bernhard85}.          

\section{Conclusions and outlook}
In this work we have presented an implementation of charge constrained 
density functional molecular dynamics in the plane-wave code CPMD. 
Several technical issues of CDFT were investigated such as the dependence 
of the results on the shape of the weight function used to define the 
charge constraint. Although for small donor acceptor-distances the energy 
depends strongly on the weight function, for medium to large distances
the dependence is rather small. Thus, it is appealing to use this
method for the study of long-range electron transfer problems.           
   
We have demonstrated that it is feasible to calculate electron 
transfer properties of condensed phase systems within the framework of 
CDFT-MD. As yet this is not possible with conventional density functional molecular 
dynamics simulation because for most donor-acceptor systems uncorrected GGA 
or hybrid density functionals give charge delocalized states 
for {\it any} nuclear configuration.   
Although the computational cost of CDFT-MD is higher than for standard 
Born-Oppenheimer dynamics, by a factor of about $2\!-\!3$, CDFT-MD proved to be 
a viable method for sampling charge-localized diabatic states of 
condensed phase systems. 

The reorganization free energy obtained for aqueous 
Ru$^{2+}$-Ru$^{3+}$ electron self-exchange, $1.62$ eV, is smaller
than estimates reported previously by other authors, 
$1.95$\,eV\cite{Rotzinger02} and $2.0$\,eV\cite{Bernhard85}. 
This can be partly explained by the short Ru-Ru distance 
of $5.5$\,\AA{} in our present simulation. Increase of this distance 
to $6.5$\,\AA{}\cite{Brunschwig82,Rotzinger02} will lead to an increase 
in reorganization free energy by about 0.2\,eV, thus bringing our 
value closer to previous estimates. However, present CDFT-MD 
simulations predict that the difference in Ru-O bond lengths in 
Ru$^{2+}$(H$_2$O)$_6$ and Ru$^{3+}$(H$_2$O)$_6$  
is significantly reduced in the electron transfer complex relative 
to the isolated aqua ions. Thus, estimation of the inner-sphere contribution
from ab-initio calculations\cite{Rotzinger02} or bond length 
differences\cite{Brunschwig82,Bernhard85} of the separated aqua-ions could 
lead to an overestimation in reorganization free energy, because 
the deformation of bond lengths in the ET complex are neglected. 

In the present work we have focused on the free energy contribution 
to the electron transfer rate constant. Following Van Voorhis and 
coworkers\cite{Wu06jcp} we will implement an approximate calculation
of the electronic transition matrix element. This will enable us to
predict absolute rates for electron or hole transfer in 
extended systems such as solvated donor-acceptor complexes and 
solids.

\appendix

\section{Weight function cutoff}
\label{app::wco}
The weight function defined in Eq.~\ref{eq::wi} is always finite even at 
points in space where all $\rho_i$ are zero. Yet, for numerical calculations 
we introduce for each atom species $k$ a radial cutoff $R_{ck}$ to avoid 
small numbers in the denominator. This also makes the real space charge
integration more efficient. The effect of the cutoff can be formally described 
by a Heaviside step function $\theta$. Thus, in Eq.~\ref{eq::wi} we 
make the following replacement,
\begin{equation}
\rho_i(\vert\mathbf{r}-\mathbf{R}_i\vert) \rightarrow
\rho_i(\vert\mathbf{r}-\mathbf{R}_i\vert)
\theta(R_{ck}-\vert\mathbf{r}-\mathbf{R}_i\vert) 
\label{eq::cutoff}
\end{equation}
where $\mathbf{R}_i$ 
is the position of atom $i$. $R_{ck}$ is chosen such that the total reference 
density of species $k$ is smaller than $10^{-6}$e, unless stated otherwise.  

\section{Constraint forces}
\label{app::cforce}
The additional force on atom $i$ due to the constraint term on the right hand side of 
Eq.~\ref{eq::Wfunc} is given by
\begin{equation}
\mathbf{F}_{\text{c}i}= - V \int \rho(\mathbf{r})
\frac{\partial w(\mathbf{r},\mathbf{R})}{\partial \mathbf{R}_i}\;d\mathbf{r}. 
\label{eq::forceapp}
\end{equation}
For the weight function defined in Eq.~\ref{eq::w} the derivative in the 
integrand of Eq.~\ref{eq::forceapp} is given by   
\begin{equation}
\frac{\partial w(\mathbf{r},\mathbf{R})}{\partial \mathbf{R}_i}=
-\frac{\rho_i^\prime(\vert\mathbf{r}-\mathbf{R}_i\vert)}
{\sum_{j=1}^N \rho_j(\vert\mathbf{r}-\mathbf{R}_j\vert)} G_i(\mathbf{r}-\mathbf{R}_i)
\label{eq::dwdr}
\end{equation}
where 
\begin{eqnarray}
G_i(\mathbf{r}-\mathbf{R}_i) = \left\lbrace
\begin{array}{l c r}
w(\mathbf{r}-\mathbf{R}_i)-1 & & i \in D\\
w(\mathbf{r}-\mathbf{R}_i)+1 & & i \in A\\
w(\mathbf{r}-\mathbf{R}_i) & & i \not\in D\cup A\\
\end{array}\right.
\end{eqnarray}
depending on whether atom $i$ is in the group of donor (D)
or acceptor (A) atoms, or in neither of them (e.g. solvent atoms). 
The derivative of the density is given by        
\begin{eqnarray}
\rho_i^\prime(\vert\mathbf{r}-\mathbf{R}_i\vert)&=&
\frac{\partial\rho_i(\vert\mathbf{r}-\mathbf{R}_i\vert)}{\partial\mathbf{R}_i}
\nonumber\\
& = &  \frac{\partial\rho_i(\vert\mathbf{r}-\mathbf{R}_i\vert)}{\partial 
\vert\mathbf{r}-\mathbf{R}_i\vert}\frac{\mathbf{r}-\mathbf{R}_i}
{\vert\mathbf{r}-\mathbf{R}_i\vert}. \label{eq::rhoprime}\\ \nonumber
\end{eqnarray}

The radial partial derivative of $\rho_i$ is 
calculated numerically using splines. However, due to the radial 
cutoff of the density (Eq.~\ref{eq::cutoff}) the derivative 
Eq.~\ref{eq::rhoprime} has to be replaced as follows, 

\begin{eqnarray}
\rho_i^\prime(\vert\mathbf{r}-\mathbf{R}_i\vert) &\rightarrow  
& \rho_i^\prime(\vert\mathbf{r}-\mathbf{R}_i\vert)\theta
(R_{ck}-\vert\mathbf{r}-\mathbf{R}_i\vert)\nonumber\\
&+&\rho_i(\vert\mathbf{r}-\mathbf{R}_i\vert)\delta
(R_{ck}-\vert\mathbf{r}-\mathbf{R}_i\vert),
			\end{eqnarray}
where $\delta$ is the Dirac delta function. Thus, the 
constraint force, Eq.~\ref{eq::forceapp}, is composed of two terms, the force 
due to $w$ within $R_{ck}$, 
$\mathbf{F}_{ci}^{\text{inside}}$, and a surface term,
$\mathbf{F}_{ci}^{\text{surf}}$, 
\begin{equation}
\mathbf{F}_{ci}=\mathbf{F}_{ci}^{\text{inside}}+\mathbf{F}_{ci}^{\text{surf}}
\end{equation}
where

\begin{eqnarray}
\mathbf{F}_{ci}^{\text{inside}} & = & \!\!- V\!\! \int \!\! \rho(\mathbf{r})
\frac{\partial w(\mathbf{r},\mathbf{R})}{\partial \mathbf{R}_i}
\;\theta(R_{ck}-\vert\mathbf{r}-\mathbf{R}_i\vert)\;d\mathbf{r} \label{eq::finside}\\ 
\mathbf{F}_{ci}^{\text{surf}} & = & -V \rho_i(R_{ck}) R_{ck} \int  
\frac{\rho(R_{ck},\vartheta,\varphi)G(R_{ck},\vartheta,\varphi)}
{\sum \rho_i(R_{ck},\vartheta,\varphi)} \nonumber\\ 
& & \times R_{ck}
\left(\begin{array}{c}
\sin\vartheta\cos\varphi\\
\sin\vartheta\sin\varphi\\
\cos\vartheta
\end{array}
\right)\sin\vartheta \;d\vartheta d\varphi \label{eq::fsurf} 
\end{eqnarray}
The derivative in the integrand of Eq.~\ref{eq::finside} is given by 
Eqs.~\ref{eq::dwdr}-\ref{eq::rhoprime} and in Eq.~\ref{eq::fsurf} we 
have changed to spherical coordinates. The surface force Eq.~\ref{eq::fsurf} 
is integrated over a thin shell on a cartesian grid. 
$R_{ck}$ times the radial unit vector is just the position vector of a point on the 
surface $(x,y,z)$. Thus the surface element can be expressed in cartesian 
coordinates as
\begin{equation}
\sin\vartheta\;d\vartheta d\varphi = \sgn(z) \frac{y\;dx dz-x\;dy dz}{R_{ck}(R_{ck}^2-z^2)}
\end{equation}
	
\section{Prediction of the Lagrange multiplier $V$ in CDFT-MD}
		\label{app::vpred}
In a deterministic Born-Oppenheimer MD simulation the ionic positions and
momenta at any given timestep depend on the configurations at earlier 
times such that -- using a suitable algorithm -- one can calculate the 
evolution of the system in time. The Lagrange multiplier $V$ used in the 
calculation of the constrained energy functional Eq.~\ref{eq::Wfunc} 
is in principle an unknown function of all ionic positions. However, the
positions change smoothly during the dynamics. Thus, one can devise an algorithm
for prediction of $V(t)$ from the history of $V$. This should provide
a better initial guess for the search of the unknown Lagrange multiplier.
We chose a Lagrange polynomial of order $n$ for this purpose\cite{Bronstein}:
		\begin{equation}
			\label{eq::Vpred}
V(t_{k+1})=\sum_{j=k-n}^{k} V(t_j) \prod_{i=k-n,i\neq j}^{k}\frac{t_{k+1}-t_i}{t_j-t_i}
		\end{equation}
where $k$ is the index of the last timestep. In Eq.~\ref{eq::Vpred}  
information from $n$ timesteps preceding step $k$ is used to extrapolate 
the value of $V$ for timestep $k+1$. Naively, one would expect higher 
order polynomials to perform better. Yet we found for CDFT-MD 
simulation of the aqueous Ru$^{2+}$-Ru$^{3+}$ complex (section \ref{sec::ru2+-ru3+}) 
an optimum extrapolation order of $n=2$. Higher order polynomials 
led to an oscillatory behaviour of Eq.~\ref{eq::Vpred} and to poor 
initial guess values. We note that the optimal order for 
extrapolation of $V$ depends very much on the system under consideration. 
Therefore it seems advisable to calculate the optimum interpolation order for 
each system in advance from a short test-run before carrying out simulations
on a larger scale. 

\begin{acknowledgments}
This work was supported by an EPSRC First grant. J. B. acknowledges The 
Royal Society for a University Research Fellowship and a research grant.
We thank Prof. M. Sprik for helpful discussions. CDFT-MD simulations were 
carried out at the High Performance Computing Facilities ``Hector'' (Edinburgh) 
and ``Darwin'' (University of Cambridge). Less time intensive calculations 
were carried out at a local compute cluster at the Center of Computational 
Chemistry, University of Cambridge.  
\end{acknowledgments}

\bibliography{bibliography}

\newcommand{\SortNoOp}[1]{}
\begin{thebibliography}{52}
\expandafter\ifx\csname natexlab\endcsname\relax\def\natexlab#1{#1}\fi
\expandafter\ifx\csname bibnamefont\endcsname\relax
  \def\bibnamefont#1{#1}\fi
\expandafter\ifx\csname bibfnamefont\endcsname\relax
  \def\bibfnamefont#1{#1}\fi
\expandafter\ifx\csname citenamefont\endcsname\relax
  \def\citenamefont#1{#1}\fi
\expandafter\ifx\csname url\endcsname\relax
  \def\url#1{\texttt{#1}}\fi
\expandafter\ifx\csname urlprefix\endcsname\relax\def\urlprefix{URL }\fi
\providecommand{\bibinfo}[2]{#2}
\providecommand{\eprint}[2][]{\url{#2}}

\bibitem[{\citenamefont{Warshel}(1982)}]{Warshel82}
\bibinfo{author}{\bibfnamefont{A.}~\bibnamefont{Warshel}},
  \bibinfo{journal}{J.~Phys.~Chem.} \textbf{\bibinfo{volume}{86}},
  \bibinfo{pages}{2218} (\bibinfo{year}{1982}).

\bibitem[{\citenamefont{Hwang and Warshel}(1987)}]{Hwang87}
\bibinfo{author}{\bibfnamefont{J.~K.} \bibnamefont{Hwang}} \bibnamefont{and}
  \bibinfo{author}{\bibfnamefont{A.}~\bibnamefont{Warshel}},
  \bibinfo{journal}{J.~Am.~Chem.~Soc.} \textbf{\bibinfo{volume}{109}},
  \bibinfo{pages}{715} (\bibinfo{year}{1987}).

\bibitem[{\citenamefont{King and Warshel}(1990)}]{King90}
\bibinfo{author}{\bibfnamefont{G.}~\bibnamefont{King}} \bibnamefont{and}
  \bibinfo{author}{\bibfnamefont{A.}~\bibnamefont{Warshel}},
  \bibinfo{journal}{J.~Chem.~Phys.} \textbf{\bibinfo{volume}{93}},
  \bibinfo{pages}{8682} (\bibinfo{year}{1990}).

\bibitem[{\citenamefont{Kuharski et~al.}(1988)\citenamefont{Kuharski, Bader,
  Chandler, Sprik, Klein, and Impey}}]{Kuharski88}
\bibinfo{author}{\bibfnamefont{R.~A.} \bibnamefont{Kuharski}},
  \bibinfo{author}{\bibfnamefont{J.~S.} \bibnamefont{Bader}},
  \bibinfo{author}{\bibfnamefont{D.}~\bibnamefont{Chandler}},
  \bibinfo{author}{\bibfnamefont{M.}~\bibnamefont{Sprik}},
  \bibinfo{author}{\bibfnamefont{M.~L.} \bibnamefont{Klein}}, \bibnamefont{and}
  \bibinfo{author}{\bibfnamefont{R.~W.} \bibnamefont{Impey}},
  \bibinfo{journal}{J.~Chem.~Phys.} \textbf{\bibinfo{volume}{89}},
  \bibinfo{pages}{3248} (\bibinfo{year}{1988}).

\bibitem[{\citenamefont{Bader et~al.}(1990)\citenamefont{Bader, Kuharski, and
  Chandler}}]{Bader90}
\bibinfo{author}{\bibfnamefont{J.~S.} \bibnamefont{Bader}},
  \bibinfo{author}{\bibfnamefont{R.~A.} \bibnamefont{Kuharski}},
  \bibnamefont{and} \bibinfo{author}{\bibfnamefont{D.}~\bibnamefont{Chandler}},
  \bibinfo{journal}{J.~Chem.~Phys.} \textbf{\bibinfo{volume}{93}},
  \bibinfo{pages}{230} (\bibinfo{year}{1990}).

\bibitem[{\citenamefont{Kakitani and Mataga}(1985)}]{Kakitani85_1}
\bibinfo{author}{\bibfnamefont{T.}~\bibnamefont{Kakitani}} \bibnamefont{and}
  \bibinfo{author}{\bibfnamefont{N.}~\bibnamefont{Mataga}},
  \bibinfo{journal}{J.~Phys.~Chem.} \textbf{\bibinfo{volume}{89}},
  \bibinfo{pages}{8} (\bibinfo{year}{1985}).

\bibitem[{\citenamefont{Tachiya}(1989)}]{Tachiya89}
\bibinfo{author}{\bibfnamefont{M.}~\bibnamefont{Tachiya}},
  \bibinfo{journal}{J.~Phys.~Chem.} \textbf{\bibinfo{volume}{93}},
  \bibinfo{pages}{7050} (\bibinfo{year}{1989}).

\bibitem[{\citenamefont{Carter and Hynes}(1989)}]{Carter89jpc}
\bibinfo{author}{\bibfnamefont{E.~A.} \bibnamefont{Carter}} \bibnamefont{and}
  \bibinfo{author}{\bibfnamefont{J.~T.} \bibnamefont{Hynes}},
  \bibinfo{journal}{J.~Phys.~Chem.} \textbf{\bibinfo{volume}{93}},
  \bibinfo{pages}{2184} (\bibinfo{year}{1989}).

\bibitem[{\citenamefont{Marcus}(1956)}]{Marcus56_1}
\bibinfo{author}{\bibfnamefont{R.~A.} \bibnamefont{Marcus}},
  \bibinfo{journal}{J.~Chem.~Phys.} \textbf{\bibinfo{volume}{24}},
  \bibinfo{pages}{966} (\bibinfo{year}{1956}).

\bibitem[{\citenamefont{Sit et~al.}(2006)\citenamefont{Sit, Cococcioni, and
  Marzari}}]{Sit06}
\bibinfo{author}{\bibfnamefont{P.~H.-L.} \bibnamefont{Sit}},
  \bibinfo{author}{\bibfnamefont{M.}~\bibnamefont{Cococcioni}},
  \bibnamefont{and} \bibinfo{author}{\bibfnamefont{N.}~\bibnamefont{Marzari}},
  \bibinfo{journal}{Phys.~Rev.~Lett.} \textbf{\bibinfo{volume}{97}},
  \bibinfo{pages}{028303} (\bibinfo{year}{2006}).

\bibitem[{\citenamefont{Zhang and Yang}(1998)}]{Zhang98}
\bibinfo{author}{\bibfnamefont{Y.}~\bibnamefont{Zhang}} \bibnamefont{and}
  \bibinfo{author}{\bibfnamefont{W.}~\bibnamefont{Yang}},
  \bibinfo{journal}{J.~Chem.~Phys.} \textbf{\bibinfo{volume}{109}},
  \bibinfo{pages}{2604} (\bibinfo{year}{1998}).

\bibitem[{\citenamefont{Mori-Sanchez
  et~al.}(2006{\natexlab{a}})\citenamefont{Mori-Sanchez, Cohen, and
  Yang}}]{Mori-Sanchez06_2}
\bibinfo{author}{\bibfnamefont{P.}~\bibnamefont{Mori-Sanchez}},
  \bibinfo{author}{\bibfnamefont{A.~J.} \bibnamefont{Cohen}}, \bibnamefont{and}
  \bibinfo{author}{\bibfnamefont{W.}~\bibnamefont{Yang}},
  \bibinfo{journal}{J.~Chem.~Phys.} \textbf{\bibinfo{volume}{125}},
  \bibinfo{pages}{201102} (\bibinfo{year}{2006}{\natexlab{a}}).

\bibitem[{\citenamefont{Mori-Sanchez
  et~al.}(2006{\natexlab{b}})\citenamefont{Mori-Sanchez, Cohen, and
  Yang}}]{Mori-Sanchez06_1}
\bibinfo{author}{\bibfnamefont{P.}~\bibnamefont{Mori-Sanchez}},
  \bibinfo{author}{\bibfnamefont{A.~J.} \bibnamefont{Cohen}}, \bibnamefont{and}
  \bibinfo{author}{\bibfnamefont{W.}~\bibnamefont{Yang}},
  \bibinfo{journal}{J.~Chem.~Phys.} \textbf{\bibinfo{volume}{124}},
  \bibinfo{pages}{091102} (\bibinfo{year}{2006}{\natexlab{b}}).

\bibitem[{\citenamefont{Cohen et~al.}(2007)\citenamefont{Cohen, Mori-Sanchez,
  and Yang}}]{Cohen07jcp1}
\bibinfo{author}{\bibfnamefont{A.~J.} \bibnamefont{Cohen}},
  \bibinfo{author}{\bibfnamefont{P.}~\bibnamefont{Mori-Sanchez}},
  \bibnamefont{and} \bibinfo{author}{\bibfnamefont{W.}~\bibnamefont{Yang}},
  \bibinfo{journal}{J.~Chem.~Phys.} \textbf{\bibinfo{volume}{126}},
  \bibinfo{pages}{191109} (\bibinfo{year}{2007}).

\bibitem[{\citenamefont{Perdew and Zunger}(1981)}]{Perdew81}
\bibinfo{author}{\bibfnamefont{J.~P.} \bibnamefont{Perdew}} \bibnamefont{and}
  \bibinfo{author}{\bibfnamefont{A.}~\bibnamefont{Zunger}},
  \bibinfo{journal}{Phys.~Rev.~B} \textbf{\bibinfo{volume}{23}},
  \bibinfo{pages}{5048} (\bibinfo{year}{1981}).

\bibitem[{\citenamefont{Tavernelli}(2007)}]{Tavernelli07}
\bibinfo{author}{\bibfnamefont{I.}~\bibnamefont{Tavernelli}},
  \bibinfo{journal}{J.~Phys.~Chem.~A} \textbf{\bibinfo{volume}{111}},
  \bibinfo{pages}{13528} (\bibinfo{year}{2007}).

\bibitem[{\citenamefont{Harrison}(1987)}]{Harrison87}
\bibinfo{author}{\bibfnamefont{J.~G.} \bibnamefont{Harrison}},
  \bibinfo{journal}{J.~Chem.~Phys.} \textbf{\bibinfo{volume}{86}},
  \bibinfo{pages}{2849} (\bibinfo{year}{1987}).

\bibitem[{\citenamefont{d'Avezac et~al.}(2005)\citenamefont{d'Avezac, Calandra,
  and Mauri}}]{d'Avezac05}
\bibinfo{author}{\bibfnamefont{M.}~\bibnamefont{d'Avezac}},
  \bibinfo{author}{\bibfnamefont{M.}~\bibnamefont{Calandra}}, \bibnamefont{and}
  \bibinfo{author}{\bibfnamefont{F.}~\bibnamefont{Mauri}},
  \bibinfo{journal}{Phys.~Rev.~B} \textbf{\bibinfo{volume}{71}},
  \bibinfo{pages}{205210} (\bibinfo{year}{2005}).

\bibitem[{\citenamefont{VandeVondele and Sprik}(2005)}]{VandeVondele05pccp}
\bibinfo{author}{\bibfnamefont{J.}~\bibnamefont{VandeVondele}}
  \bibnamefont{and} \bibinfo{author}{\bibfnamefont{M.}~\bibnamefont{Sprik}},
  \bibinfo{journal}{Phys.~Chem.~Chem.~Phys} \textbf{\bibinfo{volume}{7}},
  \bibinfo{pages}{1363} (\bibinfo{year}{2005}).

\bibitem[{\citenamefont{Liechtenstein et~al.}(1995)\citenamefont{Liechtenstein,
  Anisimov, and Zaanen}}]{Liechtenstein95}
\bibinfo{author}{\bibfnamefont{A.~I.} \bibnamefont{Liechtenstein}},
  \bibinfo{author}{\bibfnamefont{V.~I.} \bibnamefont{Anisimov}},
  \bibnamefont{and} \bibinfo{author}{\bibfnamefont{J.}~\bibnamefont{Zaanen}},
  \bibinfo{journal}{Phys.~Rev.~B} \textbf{\bibinfo{volume}{52}},
  \bibinfo{pages}{R5467} (\bibinfo{year}{1995}).

\bibitem[{\citenamefont{Migliore et~al.}(2009)\citenamefont{Migliore, Sit, and
  Klein}}]{Migliore09}
\bibinfo{author}{\bibfnamefont{A.}~\bibnamefont{Migliore}},
  \bibinfo{author}{\bibfnamefont{P.~H.-L.} \bibnamefont{Sit}},
  \bibnamefont{and} \bibinfo{author}{\bibfnamefont{M.~L.} \bibnamefont{Klein}},
  \bibinfo{journal}{J.~Chem.~Theory.~Comput.} \textbf{\bibinfo{volume}{5}},
  \bibinfo{pages}{307} (\bibinfo{year}{2009}).

\bibitem[{\citenamefont{Deskins and Dupuis}(2009)}]{Deskins09}
\bibinfo{author}{\bibfnamefont{N.~A.} \bibnamefont{Deskins}} \bibnamefont{and}
  \bibinfo{author}{\bibfnamefont{M.}~\bibnamefont{Dupuis}},
  \bibinfo{journal}{J.~Phys.~Chem.~C} \textbf{\bibinfo{volume}{113}},
  \bibinfo{pages}{346} (\bibinfo{year}{2009}).

\bibitem[{\citenamefont{Dederichs et~al.}(1984)\citenamefont{Dederichs,
  Bl{\"u}gel, Zeller, and Akai}}]{Dederichs84}
\bibinfo{author}{\bibfnamefont{P.~H.} \bibnamefont{Dederichs}},
  \bibinfo{author}{\bibfnamefont{S.}~\bibnamefont{Bl{\"u}gel}},
  \bibinfo{author}{\bibfnamefont{R.}~\bibnamefont{Zeller}}, \bibnamefont{and}
  \bibinfo{author}{\bibfnamefont{H.}~\bibnamefont{Akai}},
  \bibinfo{journal}{Phys.~Rev.~Lett.} \textbf{\bibinfo{volume}{53}},
  \bibinfo{pages}{2512} (\bibinfo{year}{1984}).

\bibitem[{\citenamefont{Akai et~al.}(1986)\citenamefont{Akai, Bl{\"u}gel,
  Zeller, and Dederichs}}]{Akai86}
\bibinfo{author}{\bibfnamefont{H.}~\bibnamefont{Akai}},
  \bibinfo{author}{\bibfnamefont{S.}~\bibnamefont{Bl{\"u}gel}},
  \bibinfo{author}{\bibfnamefont{R.}~\bibnamefont{Zeller}}, \bibnamefont{and}
  \bibinfo{author}{\bibfnamefont{P.~H.} \bibnamefont{Dederichs}},
  \bibinfo{journal}{Phys.~Rev.~Lett.} \textbf{\bibinfo{volume}{56}},
  \bibinfo{pages}{2407} (\bibinfo{year}{1986}).

\bibitem[{\citenamefont{Wu and {Van Voorhis}}(2005)}]{Wu05}
\bibinfo{author}{\bibfnamefont{Q.}~\bibnamefont{Wu}} \bibnamefont{and}
  \bibinfo{author}{\bibfnamefont{T.}~\bibnamefont{{Van Voorhis}}},
  \bibinfo{journal}{Phys.~Rev.~A} \textbf{\bibinfo{volume}{72}},
  \bibinfo{pages}{024502} (\bibinfo{year}{2005}).

\bibitem[{\citenamefont{Wu and {Van Voorhis}}(2006{\natexlab{a}})}]{Wu06jcp}
\bibinfo{author}{\bibfnamefont{Q.}~\bibnamefont{Wu}} \bibnamefont{and}
  \bibinfo{author}{\bibfnamefont{T.}~\bibnamefont{{Van Voorhis}}},
  \bibinfo{journal}{J.~Chem.~Phys.} \textbf{\bibinfo{volume}{125}},
  \bibinfo{pages}{164105} (\bibinfo{year}{2006}{\natexlab{a}}).

\bibitem[{\citenamefont{Wu and {Van Voorhis}}(2006{\natexlab{b}})}]{Wu06jpca}
\bibinfo{author}{\bibfnamefont{Q.}~\bibnamefont{Wu}} \bibnamefont{and}
  \bibinfo{author}{\bibfnamefont{T.}~\bibnamefont{{Van Voorhis}}},
  \bibinfo{journal}{J.~Phys.~Chem.~A} \textbf{\bibinfo{volume}{110}},
  \bibinfo{pages}{9212} (\bibinfo{year}{2006}{\natexlab{b}}).

\bibitem[{\citenamefont{Wu and {Van Voorhis}}(2006{\natexlab{c}})}]{Wu06jctc}
\bibinfo{author}{\bibfnamefont{Q.}~\bibnamefont{Wu}} \bibnamefont{and}
  \bibinfo{author}{\bibfnamefont{T.}~\bibnamefont{{Van Voorhis}}},
  \bibinfo{journal}{J.~Chem.~Theory.~Comput.} \textbf{\bibinfo{volume}{2}},
  \bibinfo{pages}{765} (\bibinfo{year}{2006}{\natexlab{c}}).

\bibitem[{\citenamefont{Wu and {Van Voorhis}}(2007)}]{Wu07}
\bibinfo{author}{\bibfnamefont{Q.}~\bibnamefont{Wu}} \bibnamefont{and}
  \bibinfo{author}{\bibfnamefont{T.}~\bibnamefont{{Van Voorhis}}},
  \bibinfo{journal}{J.~Chem.~Phys.} \textbf{\bibinfo{volume}{127}},
  \bibinfo{pages}{164119} (\bibinfo{year}{2007}).

\bibitem[{\citenamefont{Wu et~al.}(2009)\citenamefont{Wu, Kaduk, and {Van
  Voorhis}}}]{Wu09}
\bibinfo{author}{\bibfnamefont{Q.}~\bibnamefont{Wu}},
  \bibinfo{author}{\bibfnamefont{B.}~\bibnamefont{Kaduk}}, \bibnamefont{and}
  \bibinfo{author}{\bibfnamefont{T.}~\bibnamefont{{Van Voorhis}}},
  \bibinfo{journal}{J.~Chem.~Phys.} \textbf{\bibinfo{volume}{130}},
  \bibinfo{pages}{034109} (\bibinfo{year}{2009}).

\bibitem[{\citenamefont{Behler et~al.}(2005)\citenamefont{Behler, Delley,
  Lorenz, Reuter, and Scheffler}}]{Behler05}
\bibinfo{author}{\bibfnamefont{J.}~\bibnamefont{Behler}},
  \bibinfo{author}{\bibfnamefont{B.}~\bibnamefont{Delley}},
  \bibinfo{author}{\bibfnamefont{S.}~\bibnamefont{Lorenz}},
  \bibinfo{author}{\bibfnamefont{K.}~\bibnamefont{Reuter}}, \bibnamefont{and}
  \bibinfo{author}{\bibfnamefont{M.}~\bibnamefont{Scheffler}},
  \bibinfo{journal}{Phys.~Rev.~Lett.} \textbf{\bibinfo{volume}{94}},
  \bibinfo{pages}{036104} (\bibinfo{year}{2005}).

\bibitem[{\citenamefont{Behler et~al.}(2007)\citenamefont{Behler, Delley,
  Reuter, and Scheffler}}]{Behler07}
\bibinfo{author}{\bibfnamefont{J.}~\bibnamefont{Behler}},
  \bibinfo{author}{\bibfnamefont{B.}~\bibnamefont{Delley}},
  \bibinfo{author}{\bibfnamefont{K.}~\bibnamefont{Reuter}}, \bibnamefont{and}
  \bibinfo{author}{\bibfnamefont{M.}~\bibnamefont{Scheffler}},
  \bibinfo{journal}{Phys.~Rev.~B} \textbf{\bibinfo{volume}{75}},
  \bibinfo{pages}{115409} (\bibinfo{year}{2007}).

\bibitem[{\citenamefont{Schmidt et~al.}(2008)\citenamefont{Schmidt, Shenvi, and
  Tully}}]{Schmidt08}
\bibinfo{author}{\bibfnamefont{J.~R.} \bibnamefont{Schmidt}},
  \bibinfo{author}{\bibfnamefont{N.}~\bibnamefont{Shenvi}}, \bibnamefont{and}
  \bibinfo{author}{\bibfnamefont{J.~C.} \bibnamefont{Tully}},
  \bibinfo{journal}{J.~Chem.~Phys.} \textbf{\bibinfo{volume}{129}},
  \bibinfo{pages}{114110} (\bibinfo{year}{2008}).

\bibitem[{cpm()}]{cpmd}
\bibinfo{howpublished}{CPMD Version 3.13.2, The CPMD consortium, {\tt
  http://www.cpmd.org}, MPI f{\"u}r Festk{\"o}rperforschung and the IBM Zurich
  Research Laboratory 2009}.

\bibitem[{\citenamefont{Sulpizi et~al.}(2005)\citenamefont{Sulpizi,
  Rothlisberger, and Laio}}]{Sulpizi05}
\bibinfo{author}{\bibfnamefont{M.}~\bibnamefont{Sulpizi}},
  \bibinfo{author}{\bibfnamefont{U.}~\bibnamefont{Rothlisberger}},
  \bibnamefont{and} \bibinfo{author}{\bibfnamefont{A.}~\bibnamefont{Laio}},
  \bibinfo{journal}{Journal of Theoretical and Computational Chemistry}
  \textbf{\bibinfo{volume}{4}}, \bibinfo{pages}{985} (\bibinfo{year}{2005}).

\bibitem[{\citenamefont{Hirshfeld}(1977)}]{Hirshfeld77}
\bibinfo{author}{\bibfnamefont{F.~L.} \bibnamefont{Hirshfeld}},
  \bibinfo{journal}{Theor.~Chem.~Acc.} \textbf{\bibinfo{volume}{44}},
  \bibinfo{pages}{129} (\bibinfo{year}{1977}).

\bibitem[{\citenamefont{Tateyama et~al.}(2005)\citenamefont{Tateyama,
  Blumberger, Sprik, and Tavernelli}}]{Tateyama05}
\bibinfo{author}{\bibfnamefont{Y.}~\bibnamefont{Tateyama}},
  \bibinfo{author}{\bibfnamefont{J.}~\bibnamefont{Blumberger}},
  \bibinfo{author}{\bibfnamefont{M.}~\bibnamefont{Sprik}}, \bibnamefont{and}
  \bibinfo{author}{\bibfnamefont{I.}~\bibnamefont{Tavernelli}},
  \bibinfo{journal}{J.~Chem.~Phys.} \textbf{\bibinfo{volume}{122}},
  \bibinfo{pages}{234505} (\bibinfo{year}{2005}).

\bibitem[{\citenamefont{Blumberger et~al.}(2006)\citenamefont{Blumberger,
  Tavernelli, Klein, and Sprik}}]{Blumberger06jcp}
\bibinfo{author}{\bibfnamefont{J.}~\bibnamefont{Blumberger}},
  \bibinfo{author}{\bibfnamefont{I.}~\bibnamefont{Tavernelli}},
  \bibinfo{author}{\bibfnamefont{M.~L.} \bibnamefont{Klein}}, \bibnamefont{and}
  \bibinfo{author}{\bibfnamefont{M.}~\bibnamefont{Sprik}},
  \bibinfo{journal}{J.~Chem.~Phys.} \textbf{\bibinfo{volume}{124}},
  \bibinfo{pages}{64507} (\bibinfo{year}{2006}).

\bibitem[{\citenamefont{Blumberger}(2008)}]{Blumberger08jacs}
\bibinfo{author}{\bibfnamefont{J.}~\bibnamefont{Blumberger}},
  \bibinfo{journal}{J. Am. Chem. Soc.} \textbf{\bibinfo{volume}{130}},
  \bibinfo{pages}{16065} (\bibinfo{year}{2008}).

\bibitem[{\citenamefont{Becke}(1988)}]{Becke88}
\bibinfo{author}{\bibfnamefont{A.~D.} \bibnamefont{Becke}},
  \bibinfo{journal}{Phys.~Rev.~A} \textbf{\bibinfo{volume}{38}},
  \bibinfo{pages}{3098} (\bibinfo{year}{1988}).

\bibitem[{\citenamefont{Lee et~al.}(1988)\citenamefont{Lee, Yang, and
  Parr}}]{Lee88}
\bibinfo{author}{\bibfnamefont{C.}~\bibnamefont{Lee}},
  \bibinfo{author}{\bibfnamefont{W.}~\bibnamefont{Yang}}, \bibnamefont{and}
  \bibinfo{author}{\bibfnamefont{R.}~\bibnamefont{Parr}},
  \bibinfo{journal}{Phys.~Rev.~B} \textbf{\bibinfo{volume}{37}},
  \bibinfo{pages}{785} (\bibinfo{year}{1988}).

\bibitem[{\citenamefont{Troullier and Martins}(1991)}]{Troullier91}
\bibinfo{author}{\bibfnamefont{N.}~\bibnamefont{Troullier}} \bibnamefont{and}
  \bibinfo{author}{\bibfnamefont{J.}~\bibnamefont{Martins}},
  \bibinfo{journal}{Phys.~Rev.~B} \textbf{\bibinfo{volume}{43}},
  \bibinfo{pages}{1993} (\bibinfo{year}{1991}).

\bibitem[{\citenamefont{Blumberger and Lamoureux}(2008)}]{Blumberger08mp}
\bibinfo{author}{\bibfnamefont{J.}~\bibnamefont{Blumberger}} \bibnamefont{and}
  \bibinfo{author}{\bibfnamefont{G.}~\bibnamefont{Lamoureux}},
  \bibinfo{journal}{Mol.~Phys.} \textbf{\bibinfo{volume}{106}},
  \bibinfo{pages}{1597} (\bibinfo{year}{2008}).

\bibitem[{\citenamefont{Andersen}(1983)}]{Andersen83}
\bibinfo{author}{\bibfnamefont{H.~C.} \bibnamefont{Andersen}},
  \bibinfo{journal}{J.~Comput.~Phys.} \textbf{\bibinfo{volume}{52}},
  \bibinfo{pages}{24} (\bibinfo{year}{1983}).

\bibitem[{\citenamefont{Martyna et~al.}(1992)\citenamefont{Martyna, Klein, and
  Tuckerman}}]{Martyna92}
\bibinfo{author}{\bibfnamefont{G.~J.} \bibnamefont{Martyna}},
  \bibinfo{author}{\bibfnamefont{M.~L.} \bibnamefont{Klein}}, \bibnamefont{and}
  \bibinfo{author}{\bibfnamefont{M.}~\bibnamefont{Tuckerman}},
  \bibinfo{journal}{J.~Chem.~Phys.} \textbf{\bibinfo{volume}{97}},
  \bibinfo{pages}{2635} (\bibinfo{year}{1992}).

\bibitem[{\citenamefont{Pieniazek et~al.}(2007)\citenamefont{Pieniazek,
  Arnstein, Bradforth, Krylov, and Sherrill}}]{Pieniazek07}
\bibinfo{author}{\bibfnamefont{P.~A.} \bibnamefont{Pieniazek}},
  \bibinfo{author}{\bibfnamefont{S.~A.} \bibnamefont{Arnstein}},
  \bibinfo{author}{\bibfnamefont{S.~E.} \bibnamefont{Bradforth}},
  \bibinfo{author}{\bibfnamefont{A.~I.} \bibnamefont{Krylov}},
  \bibnamefont{and} \bibinfo{author}{\bibfnamefont{C.~D.}
  \bibnamefont{Sherrill}}, \bibinfo{journal}{J.~Chem.~Phys.}
  \textbf{\bibinfo{volume}{127}}, \bibinfo{pages}{164110}
  (\bibinfo{year}{2007}).

\bibitem[{\citenamefont{Blumberger and Sprik}(2006)}]{Blumberger06tca}
\bibinfo{author}{\bibfnamefont{J.}~\bibnamefont{Blumberger}} \bibnamefont{and}
  \bibinfo{author}{\bibfnamefont{M.}~\bibnamefont{Sprik}},
  \bibinfo{journal}{Theor.~Chem.~Acc.} \textbf{\bibinfo{volume}{115}},
  \bibinfo{pages}{113} (\bibinfo{year}{2006}).

\bibitem[{\citenamefont{Blumberger and Sprik}(2005)}]{Blumberger05jpcb}
\bibinfo{author}{\bibfnamefont{J.}~\bibnamefont{Blumberger}} \bibnamefont{and}
  \bibinfo{author}{\bibfnamefont{M.}~\bibnamefont{Sprik}}, \bibinfo{journal}{J.
  Phys. Chem. B} \textbf{\bibinfo{volume}{109}}, \bibinfo{pages}{6793}
  (\bibinfo{year}{2005}).

\bibitem[{\citenamefont{Rotzinger}(2002)}]{Rotzinger02}
\bibinfo{author}{\bibfnamefont{F.~P.} \bibnamefont{Rotzinger}},
  \bibinfo{journal}{J.~Chem.~Soc.~Dalton~Trans.} p. \bibinfo{pages}{719}
  (\bibinfo{year}{2002}).

\bibitem[{\citenamefont{Bernhard et~al.}(1985)\citenamefont{Bernhard, Helm,
  Ludi, and Merbach}}]{Bernhard85}
\bibinfo{author}{\bibfnamefont{P.}~\bibnamefont{Bernhard}},
  \bibinfo{author}{\bibfnamefont{L.}~\bibnamefont{Helm}},
  \bibinfo{author}{\bibfnamefont{A.}~\bibnamefont{Ludi}}, \bibnamefont{and}
  \bibinfo{author}{\bibfnamefont{A.~E.} \bibnamefont{Merbach}},
  \bibinfo{journal}{J.~Am.~Chem.~Soc.} \textbf{\bibinfo{volume}{107}},
  \bibinfo{pages}{312} (\bibinfo{year}{1985}).

\bibitem[{\citenamefont{Brunschwig et~al.}(1982)\citenamefont{Brunschwig,
  Creutz, McCartney, Sham, and Sutin}}]{Brunschwig82}
\bibinfo{author}{\bibfnamefont{B.~S.} \bibnamefont{Brunschwig}},
  \bibinfo{author}{\bibfnamefont{C.}~\bibnamefont{Creutz}},
  \bibinfo{author}{\bibfnamefont{D.~H.} \bibnamefont{McCartney}},
  \bibinfo{author}{\bibfnamefont{T.-K.} \bibnamefont{Sham}}, \bibnamefont{and}
  \bibinfo{author}{\bibfnamefont{N.}~\bibnamefont{Sutin}},
  \bibinfo{journal}{Faraday~Discuss.~Chem.~Soc.} \textbf{\bibinfo{volume}{74}},
  \bibinfo{pages}{113} (\bibinfo{year}{1982}).

\bibitem[{\citenamefont{Bronstein et~al.}(1999)\citenamefont{Bronstein,
  Semendjajew, Musiol, and M{\"u}hlig}}]{Bronstein}
\bibinfo{author}{\bibfnamefont{I.~G.} \bibnamefont{Bronstein}},
  \bibinfo{author}{\bibfnamefont{K.~A.} \bibnamefont{Semendjajew}},
  \bibinfo{author}{\bibfnamefont{G.}~\bibnamefont{Musiol}}, \bibnamefont{and}
  \bibinfo{author}{\bibfnamefont{H.}~\bibnamefont{M{\"u}hlig}},
  \emph{\bibinfo{title}{Taschenbuch der Mathematik}}
  (\bibinfo{publisher}{Verlag Harri Deutsch}, \bibinfo{year}{1999}),
  \bibinfo{edition}{4th} ed.

\end{thebibliography}
\end{document}